\documentclass[aps,amsmath,amssymb,superscriptaddress,showpacs,floatfix,twocolumn, pra,reprint]{revtex4-1}
\usepackage{graphicx,xspace,epsfig,float,multirow}

\usepackage[T1]{fontenc}
\usepackage[latin2]{inputenc}
\usepackage{color}
\usepackage{etoolbox}
\apptocmd{\sloppy}{\hbadness 10000\relax}{}{}

\newcommand{\ket}[1]{| #1 \rangle}
\newcommand{\bra}[1]{\langle #1 |}

\newcommand{\ketbra}[2]{| #1 \rangle \langle #2 |}

\newcommand{\non}{\nonumber}
\newcommand{\I}{{\rm{i}}}
\newcommand{\D}{{\rm{d}}}

\newcommand{\nn}{\nonumber}
\newcommand{\Heff}{\hat{H}_{\rm eff}}
\newcommand{\nopone}{\hat{n}_1}
\newcommand{\noptwo}{\hat{n}_2}
\newcommand{\nopi}{\hat{n}_i}

\newcommand{\aopone}{\hat{a}_1}
\newcommand{\aoptwo}{\hat{a}_2}
\newcommand{\aopi}{\hat{a}_i}

\newcommand{\lt}{\left(}
\newcommand{\rt}{\right)}

\newcommand{\be}{\begin{equation}}
\newcommand{\ee}{\end{equation}}
\newcommand{\ba}{\begin{eqnarray}}
\newcommand{\ea}{\end{eqnarray}}

\def\proba{{\rm I\kern -.18em P}}

\newcommand{\sinc}{\operatorname{sinc}}

\newcommand{\ie}{i.e.}

\begin{document}

\title{Macroscopic superpositions in Bose-Josephson junctions: controlling
 decoherence due to atom losses}
 \author{K. Pawlowski}
 \affiliation{Center for Theoretical Physics PAN, Al. Lotnik\'ow 32/46, 02-668
Warsaw, Poland}
 \affiliation{5. Physikalisches Institut, Universit\"at Stuttgart,
 Pfaffenwaldring 57, D-70569 Stuttgart, Germany}
 \affiliation{Laboratoire Kastler Brossel, Ecole Normale Sup\'erieure, 45, rue d'Ulm,  75231 Paris, France}
 \author{D. Spehner}
 \email{Dominique.Spehner@ujf-grenoble.fr}
 \affiliation{Universit\'e Grenoble 1 and CNRS, Institut Fourier  UMR5582,
 B.P. 74, 38402 Saint Martin d'H\`eres, France}
 \affiliation{Universit\'e Grenoble 1 and CNRS, Laboratoire de Physique et
Mod\'elisation des
 Milieux Condens\'es UMR5493, B.P. 166, 38042 Grenoble, France}
 \author{A. Minguzzi}
 \affiliation{Universit\'e Grenoble 1 and CNRS, Laboratoire de Physique et
Mod\'elisation des
 Milieux Condens\'es UMR5493, B.P. 166, 38042 Grenoble, France}
 \author{G. Ferrini}
 \affiliation{Laboratoire Kastler Brossel, Universit\'e Pierre et Marie Curie,
4 place de Jussieu,
 75005 Paris, France}

\date{\today}

\begin{abstract}
We study how macroscopic superpositions of coherent states produced by the
nondissipative dynamics of binary mixtures of ultracold atoms are affected by
atom losses. We identify different decoherence scenarios for symmetric or
asymmetric loss rates and interaction energies in the two modes. 
In the symmetric case the quantum coherence in the superposition is 
lost after a single loss event.  
By tuning appropriately the energies we show that the superposition can be protected, leading to  quantum correlations useful for atom interferometry even after many loss events.
\end{abstract}


\maketitle

\section{Introduction}

In large systems, macroscopic superpositions of quantum states 
 have extremely small decoherence times, making them  impossible to observe~\cite{Giulini}.
 The largest 
 superpositions  generated so far are superpositions of coherent states (CSs)
of a cavity field,
and  their progressive transformation into statistical mixtures 
as time evolves has been observed~\cite{Haroche}. 
The first proposal to generate a macroscopic superposition of CSs (MSCSs) with light 
consisted of sending photons through
a medium presenting a strong Kerr nonlinearity~\cite{yurke1986}. In such media, the
dynamical phases of Fock states
are nonlinear in the photon number, thus the phase of an initial 
CS is split. In metastable vapors of 
 ultracold bosonic atoms, interactions between atoms lead to similar nonlinearities,
whose strength can be tuned experimentally by using Feshbach resonances~\cite{fano1961,feshbach1958}.
By trapping optically the condensed atoms in a double-well potential one realizes 
an external Bose-Josephson junction (BJJ); an internal BJJ is obtained by
trapping in a single well atoms in two distinct hyperfine states coupled by a resonant field.  
In analogy with light, the nonlinear dynamics   
generates MSCSs after a sudden quench to zero of the tunnel amplitude (for external BJJs) or a switch off of
the coupling field (for internal BJJs)~\cite{piazza2008,ferrini2008}. 
Whereas MSCSs of light are presumably destroyed after a single photon loss~\cite{Haroche}, the situation 
is less clear for atoms. 
Up to now only squeezed states -- which are produced at earlier  times than MSCSs --
have been observed  in BJJs with 
a few hundreds of atoms~\cite{esteve2008, riedel2010, gross2010}.
 In order to know if future experiments  could produce MSCSs 
in BJJs, it is desirable to study how robust they are with respect to decoherence.   

Decoherence effects due to  atom losses  and phase noise  on squeezed states have been
analyzed in detail \cite{sinatra1998, sinatra2012, pawlowskiBackground}.
For what concerns MSCSs, only decoherence caused by  noise due to photon 
scattering \cite{huang2006} and 
phase noise \cite{ferrini2010a} has been studied so far. 
Under current experimental conditions the first noise is
negligible and the second one can be reduced by using spin echo~\cite{gross2010,gross2010dissJPB}; in contrast,
atom losses are  unavoidable.

In this article, we investigate whether a macroscopic superposition can be formed in a BJJ even
 in the presence of atom losses. We are primarily interested in internal Bose-Josephson junctions, 
such as those studied experimentally in~\cite{riedel2010, gross2010}. 
We focus on two-body losses, due to
scattering of two atoms in the magnetic trap which changes their spin and gives them enough kinetic energy 
to be ejected from the trap.  These loss processes are particularly detrimental in the 
experiments of Refs.~\cite{gross2010, riedel2010}.
Our analysis, however, also applies to one- and three-body 
losses~\cite{long_article}.
We first analyze the dynamics of a lossy BJJ from the point of view of state conditioning,  assuming that 
the total number $\hat{N}$ of condensed atoms can be measured precisely, both initially and at the MSCSs 
formation time $t_q$. We study how much coherence 
is destroyed by a single loss event occurring at a random time between $0$ and $t_q$.
We find quite different answers depending on the degree of asymmetry between the loss rates and 
interaction energies in the two modes of the junction. 
Finally, we show that for strongly asymmetric losses one can protect
the coherence of the MSCSs by suitably tuning  the interaction energies, even  after many loss events and 
in the absence of measurement of the atom number.

\vspace{1cm}

\section{Quenched dynamics in Bose-Josephson junctions in the presence of atom losses} 

We consider an internal BJJ in the quantum regime. 
Initially, $N_0$ atoms are all in the same single-particle state which is a symmetric
superposition of the two internal 
states. This corresponds to the  ground state in the regime where 
tunneling dominates interactions, described by a spin coherent state 
\begin{eqnarray} \label{eq-def_coherent_state}
& & \ket{\psi (0)}  =  \ket{N_0;\theta,\phi} 
\\
\non
& & =  
\sum_{n_1=0}^{N_0}\!\! \lt \begin{array}{c} N_0 \\ n_1 \end{array} \rt^{1/2}
\!\!\!\!\!
  \frac{\,(e^{-\I\phi} \tan (\theta/2))^{n_1}}{[1+\tan^2 (\theta/2)]^{N_0/2}} 
\ket{n_1,N_0-n_1}
\end{eqnarray}
 with $\theta=\pi/2$ and $\phi=0$. Here
$\ket{n_1,n_2}$ is the joint eigenstate of 
the number operators $\nopi$ in the mode $i=1,2$ (Fock state).
 The dynamics following a sudden quench of the coupling (tunnel energy) to zero 
is given by the two-mode Bose-Hubbard Hamiltonian~\cite{Milburn97}
\begin{equation} \label{eq-H_0}
\hat{H}_0 = \sum_{i=1,2} \lt E_i \nopi + \frac{U_i}{2} \nopi (\nopi -1) \rt
+ U_{12} \nopone \noptwo ,
\end{equation}
where  $E_i$ and $U_i$ are the internal energy and the 
interaction energy  between two atoms in the same mode $i$, respectively, and $U_{12}$ is the  
inter-mode interaction energy. 
Setting $\noptwo=N_0-\nopone$, the interactions  in Eq.~(\ref{eq-H_0}) sum up to a 
non-linear term $\chi \nopone^2$,  with $\chi= (U_1+U_2-2 U_{12})/2$.
The time-evolved state at time $t_q=\pi/|\chi q|$
is a superposition of CSs,
$\ket{\psi^{(0)} (t_q)} = e^{-\I t_q \hat{H}_0} \ket{\psi (0)}
=\sum_{k=0}^{q-1} c_k \ket{N_0;\frac{\pi}{2},\phi_k}$,
 with $\phi_{k+1}-\phi_k =2\pi/q$ and
$|c_k | = q^{-1/2}$~\cite{yurke1986}.

In the presence of two-body losses the  markovian master equation for
the density matrix $ \hat{\rho}(t)$ of the condensed atoms reads~\cite{anglin1997,jack2002} 
 (setting $\hbar=1$)
\begin{eqnarray} \label{eq-master_equation}
& & 
\frac{\D \hat{\rho}(t)}{\D t} 
= - \I \bigl[ \hat{H}_0 , \hat{\rho} \bigr] 
+  \sum_{i=1,2} \gamma_i 
  \Bigl( 
   \aopi^2 \, \hat{\rho} \, ({\aopi}^2)^\dagger  
\\
\nn
& & 
- \frac{1}{2} \bigl\{ \nopi ( \nopi -1) ,  \hat{\rho}  \bigr\}
  \Bigr)
 + \gamma_{12} 
\Bigl(
 \aopone \aoptwo \, \hat{\rho}  \, \aopone^\dagger \aoptwo^\dagger 
- \frac{1}{2} \bigl\{ \nopone \noptwo , \hat{\rho}  \bigr\} 
\Bigr)
\end{eqnarray}
where $\gamma_i$ and $\gamma_{12}$ are the loss rates of two atoms in the same mode $i$
and of one atom in each mode, respectively, 
and $\{ \cdot , \cdot\}$ denotes the anticommutator.
 Since Eq.~(\ref{eq-master_equation}) does not couple subspaces with different
total atom numbers $\hat{N} = \nopone + \noptwo$  and $\hat{N}=N_0$ initially,  at all times
$\hat{\rho} (t )$ has the block structure 
$\hat{\rho} (t ) = \sum_{N=0}^{N_0} w_N (t) \hat{\rho}_N (t)$, where
$w_N(t)$  is the probability to have $N$ atoms at time $t$ and
 $\hat{\rho}_N (t)$ is the  corresponding conditional state.
This decomposition is naturally accounted for by 
quantum trajectories $t \mapsto \ket{\psi_J(t)}= \ket{\widetilde{\psi}_J(t)}/\| \widetilde{\psi}_J(t) \|$, 
$\ket{\widetilde{\psi}_J(t)}$ being the unnormalized wave function when
$J$ loss events  occur at times 
$0 \leq s_1 \leq s_2 \leq \cdots \leq s_J \leq t$  in the 
channels $m_1, \cdots ,m_J$~\cite{Haroche, carmichael1991,dalibard1992},  
\begin{eqnarray}  
\label{eq-no_jump}
\ket{\widetilde{\psi}_J (t)} 
&  = &  
  e^{-\I(t-s_J)  \Heff } \hat{M}_{m_J} e^{-\I (s_J-s_{J-1})  \Heff}
\hat{M}_{m_{J-1}} \cdots 
\nonumber \\
& & 
\cdots e^{-\I \Heff (s_2-s_1) } \hat{M}_{m_1} e^{-\I s_1 \Heff } \ket{\psi (0)} .
\end{eqnarray}
For two-body losses one has three loss channels $m  = 1$, $2$, and $12$. The corresponding
jump operators are $\hat{M}_{1}= {\aopone}^2$, $\hat{M}_2 = {\aoptwo}^2$, and 
$ \hat{M}_{12}= \aopone\aoptwo$.
The dynamics between loss events is given by the effective Hamiltonian 
$\Heff = \hat{H_0} - \I \hat{D}$. The damping operator
\begin{equation} \label{eq-D}
\hat{D} =  \frac{1}{2} \Bigl( \sum_{i=1,2} \gamma_i  \nopi ( \nopi -1 ) 
+  \gamma_{12} \nopone \noptwo \Bigr)
\end{equation}
describes the
gain of information on the system resulting from the knowledge that no loss occurred.
The conditional state $\hat{\rho}_{N} (t)$ after a detection
of $N=N_0-2J$ atoms at time $t$ is obtained by
averaging $\ketbra{{\psi}_J(t)}{{\psi}_J(t)}$ over the $J$ jump times $s_K$ and channels $m_K$, 
\begin{eqnarray} \label{eq-density_matrix_subspace_N_0-j}
\non
\widetilde{\rho}_{N}(t) & \equiv & w_{N} (t) \hat{\rho}_{N} (t) = \sum_{m_1,\cdots,m_J} 
 \int_{0\leq s_1 \leq \cdots \leq s_J \leq t}\!\!\!\!\!\!\!\!\!\!\!\!  \D s_1 
\ldots \D s_J\,
\\
& & 
\times p_{m_1,\ldots,m_J}^{(t)} (s_1,\ldots , s_J ; J ) 
\ketbra{{\psi}_J(t)}{{\psi}_J(t)} \;,
\end{eqnarray}
where $
p_{m_1,\ldots,m_J}^{(t)} (s_1,\ldots , s_J ; J ) 
 =  \gamma_{m_1} \ldots \gamma_{m_J} \| {\widetilde{\psi}_J(t)} \|^2
$
is the joint distribution of the  $s_K$, $m_K$, and $J$~\cite{dalibard1992}. 
By further summing over the number of jumps $J$ one gets 
the total density matrix $\hat{\rho} (t ) = \sum_{N=0}^{N_0} w_N (t)
\hat{\rho}_N (t)$ which is a solution of Eq.~(\ref{eq-master_equation}).
In order to understand the effect of losses  we analyze separately
 each $N$-atom sectors.

\section{Conditional states and their quantum correlations}
\subsection{Density matrix in the subspace with $N_0$ atoms} 

When  no  loss occurs in the time interval $[0,t]$, from Eqs.~(\ref{eq-no_jump}) and (\ref{eq-D}) we obtain   
the unnormalized conditional state 
$\widetilde{\rho}_{N_0}^{\rm{(no\,loss)}} (t)=\ketbra{\widetilde{\psi}_{0} (t)}{\widetilde{\psi}_{0} (t)}$ 
in  the Fock basis, 
\begin{eqnarray} 
\label{eq-state_no_jump}
\non
& & \bra{n_1,n_2}  \widetilde{\rho}_{N_0}^{\rm{(no\,loss)}} (t)  \ket{n_1',n_2'}  
\\ 
& &  \hspace*{0.5cm}
=  e^{-t [d_{N_0} (n_1)+d_{N_0} (n_1')]}
 \bra{n_1,n_2} \hat{\rho}^{(0)} (t) \ket{n_1',n_2'}\,,
\end{eqnarray}
where $\hat{\rho}^{(0)} (t) = \ketbra{\psi^{(0)} (t)}{\psi^{(0)} (t)}$ 
is the lossless density matrix and
\begin{equation}
d_{N_0}(n_1)=\frac{1}{2} (\gamma_1+\gamma_2-\gamma_{12}) (n_1 - \overline{n}_1)^2
\end{equation}
up to  an irrelevant constant, with
$2 \overline{n}_1 = [\gamma_1 - \gamma_2 + N_0 ( 2 \gamma_2 - \gamma_{12})]/(\gamma_1+\gamma_2-\gamma_{12})$.
For  $N_0 \gg 1$ the  matrix elements of $\hat{\rho}^{(0)} (t)$
have Gaussian moduli peaked at $(n_1,n_1')= (N_0/2,N_0/2)$ with a width $\sim \sqrt{N_0}$. 
For symmetric loss rates $\gamma_1=\gamma_2$ and $\gamma_{12}=0$, 
this peak coincides with the center of the Gaussian damping factor
in Eq.~(\ref{eq-state_no_jump}). Thus the MSCSs formed at time $t_q$ is affected by damping when 
$\gamma_1 \gtrsim |\chi| q/N_0$. 
In contrast, for
$\gamma_2=\gamma_{12}=0$ the damping factor is centered at $(n_1,n_1')= (1/2,1/2)$ and  
its effect on the MSCSs sets
in at the much smaller rate $\gamma_1 \approx |\chi| q/N_0^2$.

\subsection{Density matrix in the subspace with $(N_0-2)$ atoms:
Tuning the energies to protect the coherence} \label{sec-N_0-2}

We focus now on trajectories having one loss event in channel $m$
at the random time $s \in [0,t]$. 
The instantaneous jump  transforms a CS into a CS, 
$\hat{M}_{m} \ket{N_0;\theta,\phi} \propto \ket{N_0-2;\theta,\phi}$.
This CS is rotated on the Bloch sphere by the evolution under the nonlinear
effective Hamiltonian $\Heff$ 
 due to changes in energy and damping before and after the jump, yielding (see Appendix~\ref{sec-app1})
 \begin{equation} 
\label{eq-wave function_one_jump_bis_bis}
\ket{{\psi}_{1} (t)}
\propto   e^{-\I t \Heff }  \ket{N_0-2; \theta_m (s) , \phi_m (s) } \;,
\end{equation}
where the
 random angles $\theta_m (s)  =  2 \arctan ( e^{- s \delta_m} )$ and
$\phi_m (s)  =  2 s \chi_m$ depend on the random loss time $s$,
the  interaction energies  
$\chi_1 =U_1-U_{12}$, $\chi_2 = - (U_2 - U_{12})$, $\chi_{12} = (U_1 - U_2)/2$, 
and the loss rate differences $\delta_{1} = 2 \gamma_1 -  \gamma_{12}$, $\delta_{2} = -(2 \gamma_2 - \gamma_{12})$,
and $\delta_{12} = \gamma_1 - \gamma_2$.   
Hence,
apart from reducing $N$ and producing damping,  atom losses are identical to external
$\theta$ and $\phi$ noises rotating the state around the $z$ axis by a complex angle 
$\phi_m+\I \ln \tan (\frac{\theta_m}{2})$~\cite{remark}.
These noises have fluctuations 
$\delta \theta_{m} \simeq  |\delta_m | \min \{  t  ,  \delta s_m \}$ and 
$\delta \phi_m  =  2 |\chi_m | \min \{ t ,   \delta s_m \}$, where 
$\delta s_m$ is the loss-time fluctuation and we assumed $\delta \theta_m \ll 1$.

%
\begin{figure}
\includegraphics[width=0.45\linewidth]{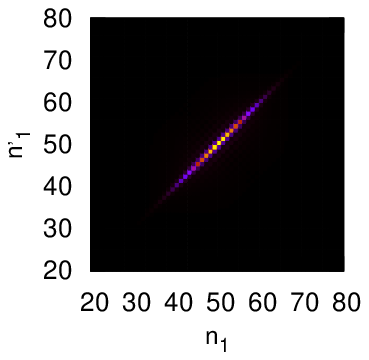}
\includegraphics[width=0.53\linewidth]{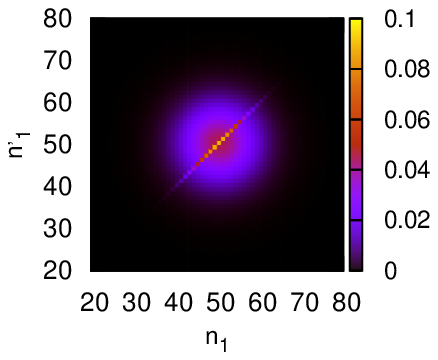}

\includegraphics[width=0.45\linewidth]{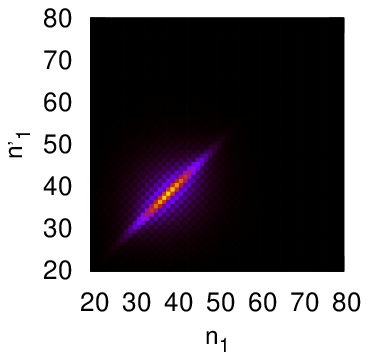}
\includegraphics[width=0.53\linewidth]{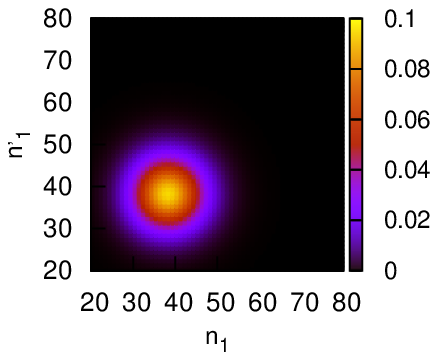}
\caption{\label{fig2} (Color online) Modulus of the density matrix 
$\hat{\rho}_{N_0-2}(t_2)$ in the subspace with
$N_0-2$ atoms
in the Fock basis at time $t_2$, obtained by an exact diagonalization of Eq.~(\ref{eq-master_equation}). 
Upper  panels: symmetric losses rates
($\gamma_1 = \gamma_2 =\chi/(200\pi)$); lower panels: asymmetric losses 
($\gamma_1= 4\chi/(300 \pi)$, $\gamma_2=0$). 
Left column: symmetric energies  ($\chi_1 = -\chi_2 =\chi$);
right column: asymmetric  energies ($\chi_1 = 2 \chi$, $\chi_2 =0$). 
Other parameters: $\gamma_{12}=E_i=0$, $N_0=100$.
}
\end{figure}

 Let us concentrate on the MSCS formation time $t=t_q$ 
and take $\gamma_{12}=0$. We first consider 
{\it weak losses $\gamma_m \lesssim q \chi/N_0$} and {\it symmetric energies $U_1=U_2$}.
In this regime the $\theta$-noise is negligible since 
$\delta \theta_m$ is much smaller than the quantum fluctuations $\sim 1/\sqrt{ N_0}$ of a CS.
In contrast, one has large $\phi$ fluctuations $\delta \phi_m = 2 \pi/q$ in the two loss channels
$m=1,2$, equal to 
the inter-component phase separation of the 
MSCSs (we use here  $\chi_{1}=-\chi_{2} = \chi$ and $\delta s_m > t_q$).
Such a strong $\phi$ noise has a relatively small effect on 
the coherences, albeit it transforms the corresponding statistical mixture 
of CSs into a mixture
of Fock states (phase relaxation)~\cite{ferrini2010a}.
Actually, in the absence of losses  in the second mode ($\gamma_2= 0$), 
the off-diagonal elements of the  conditional density matrix  
$\bra{n_1,n_2} \hat{\rho}_{N_0-2}(t_q) \ket{n_1',n_2'}$ are weakly affected 
by losses for $n_1'\neq n_1$ modulo $q$, 
as shown in Appendix~\ref{sec-app3} and in Fig.\ \ref{fig2} (lower left panel).
The global shift to values $n_1,n_1'< N_0/2$ is due to the Gaussian damping caused by
$\Heff$ in Eq.~(\ref{eq-wave function_one_jump_bis_bis}).  
For $\gamma_1=\gamma_2$, instead,   
$ \hat{\rho}_{N_0-2}(t_2)$
is almost diagonal in the Fock basis, see the upper left panel in Fig.\ref{fig2}.
This is due to a cancellation (occurring only for $q=2$)
when summing the contributions of the two channels (see Appendix~\ref{sec-app3}).
 
We now consider {\it asymmetric energies $U_1 \not= U_2$}, 
still assuming weak losses $\gamma_m \lesssim q \chi/N_0$.
In order to keep  $t_q= \pi/|\chi q|$ constant, we 
vary $\chi_{1}$, $\chi_{2}$ while fixing  $2 \chi = \chi_{1}-\chi_{2}$.
Interestingly, it is possible to  protect   one channel, say $m=1$, against $\phi$ noise  
by choosing $\chi_{1} =0$ and $\chi_{2} = -2 \chi$, at the expense of enlarging
noise in the other channel. 
Then $\delta \phi_{1}=0$  and $\delta \phi_{2}=4\pi/q$.  
If only the first channel loses atoms, the conditional state after a single loss event 
is then close to a  MSCSs with $N_0-2$ atoms, apart from the damping described in Eq.~(\ref{eq-state_no_jump}),
as seen in the lower right panel in Fig.\ \ref{fig2}. 
For symmetric losses, $\hat{\rho}_{N_0-2}(t_q)$  has also large off-diagonal 
elements in the Fock basis due to the probability $1/2$ of losing atoms in the protected channel 
(Fig.\ \ref{fig2}, upper right panel). 
Hence, the MSCSs can be protected 
by tuning the interaction energies so that $\chi_i=0$ in the mode $i$ with the highest loss rate $\gamma_i$.
For $^{87}$Rb atoms used in the experiment of Ref.~\cite{gross2010},   
a magnetic field in resonance with one of 
the Feshbach peaks for $m=1$, $2$, or $12$ must be applied in order to have a nonzero $\chi$
(actually, without magnetic field one has nearly $U_1 = U_2 = U_{12}$).
Since two-body losses are mostly
important in the upper internal level $m=2$, in order to better preserve the coherence,
$U_1$ must be tuned such that $\chi_1=2\chi \not=0$ .
Let us remark that, although our results also apply to external BJJs,
for such BJJs $U_{12}=0$ and thus one must tune  the interaction energy $U_i$ to zero  to
switch-off phase noise in the well $i$.
But the loss rate $\gamma_i$  depends on $U_i$ and this tuning actually decreases 
$\gamma_i$, so that
the protection of the MSCSs is a trivial effect. In contrast,  
for internal BJJs choosing $U_i \simeq U_{12}$ does not 
decrease the loss rates, but it diminishes decoherence in the loss channel $i$ at weak losses.

Let us now turn  to the {\it intermediate loss rate} regime. 
The $\phi$ noise decreases  when increasing $\gamma_m$
since  the loss-time fluctuations $\delta s_{m}$
decrease. 
Indeed, we find 
$\delta s_{i} \approx (2 \gamma_i + \gamma_{12})^{-1} N_0^{-1}$ ($i=1,2$) for $N_0 \gg 1$.
Physically, at increasing $\gamma_m$  
 the loss has more chance to 
occur at small times,  while for small $\gamma_m$, $s$ is
equally distributed in $[0,t]$.
Note, however,  that the probability $w_{N_0-2}(t_q)$ of losing only two atoms
decreases  by increasing $\gamma_m$. 

\subsection{Density matrix in the subspaces with $(N_0-2J)$ atoms}

 The wavefunction $\ket{{\psi}_J(t)}$ after $J>1$ jumps 
is still given by Eq.~(\ref{eq-wave function_one_jump_bis_bis}) upon replacing $N_0-2$ by $N_0-2J$ and the
angles of the CS by $\phi^{(J)}$ and  $\theta^{(J)}$ with
\begin{equation}
\phi^{(J)} \! = \! \sum_{K=1}^J \phi_{m_K} (s_K) \;,\;
\tan \Bigl(\frac{\theta^{(J)}}{2} \Bigr) \! = \! \prod_{K=1}^J \tan \Bigl(\frac{\theta_{m_K}(s_K)}{2} \Bigr)
\, ,
\end{equation}
 where
$\phi_m(s)$ and $\theta_m(s)$ are the angles corresponding to the single loss event. 
Thus the aforementioned effects  persist.
For weak symmetric losses, though, keeping
coherence by switching off phase noise in one channel is harder since
the probability that all jumps occur in that channel decreases exponentially with $J$.  
This means that for many loss events our proposal for protecting MSCSs is only efficient for 
strongly asymmetric loss rates.

\section{Total quantum correlations and sub-shot noise atom interferometry}
  
 We now show that for strongly asymmetric losses
not only the conditional states but also the quantum correlations in the full density matrix $\hat{\rho}(t_q)$
can be protected  by tuning the interactions.
We measure the amount of quantum correlations with 
the quantum Fisher information $F (\hat{\rho})$,
related to the best achievable phase precision in a Mach-Zehnder interferometer
using $\hat{\rho}$ as input state by $(\Delta \varphi)_{\text{best}}= 1/\sqrt{F(\hat{\rho})}$
\cite{braunstein1994}. Hence $F(\hat{\rho}) >  \langle \hat{N} \rangle $ implies
phase accuracy  beyond the shot noise limit 
$(\Delta \varphi )_{\rm SN} = \langle \hat{N} \rangle^{-1/2}$ \cite{HyllusPRL2010}. 
In our case 
$F_{\rm tot} (t) \equiv F [ \hat{\rho}(t)  ] = \sum_{N=0}^{N_0} w_N(t)  F_N (t)$
where $F_N (t)$  is the Fisher information of the conditional
state $\hat{\rho}_N(t)$ in the subspace with $N$ atoms. The latter is given by 
$F_N = 2\sum_{k,l}\frac{(p_k-p_l)^2}{p_k+p_l} | \langle k | \hat{J}_{{\bf n}} | l \rangle |^2$ where
   $\hat{\rho}_N | l \rangle  =  p_l | l \rangle$ and $\hat{J}_{\bf{n}}$ is the angular momentum
 operator in the direction ${\bf n}$.
We optimize $F_{\rm tot} (t)$   over all directions ${\bf n}$ 
of the interferometer.
In a lossless BJJ, the two-component superposition 
has  the highest possible value 
 $F_{\rm tot} (t_2) = N_0^2$, nearly twice larger than  that of  MSCSs with $q>2$ components~\cite{pezze09,ferrini2010}. 

\begin{figure}
\centering
\includegraphics[width=.96\columnwidth]{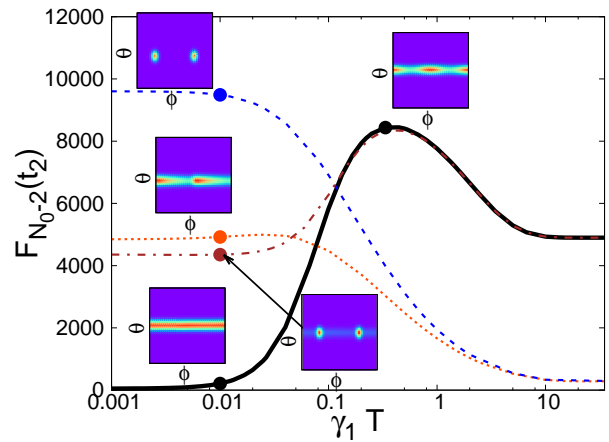}
\caption{\label{fig3} (Color online) Fisher information in the 
subspace  with $N_0-2$ atoms
as a function of the loss
rate $\gamma_1$ (in semilogarithmic scale) at time $t_2=T/4$, with $T=2\pi/\chi$.
Solid black line: $\gamma_1 = \gamma_2$,
$\chi_1 = -\chi_2 =\chi$;
  dot-dashed brown line:  $\gamma_1 = \gamma_2$, $\chi_1=0$,
$\chi_2 = - 2 \chi$;
dotted red line:  $\gamma_2 =0$, $\chi_1 = -\chi_2 = \chi$;
 dashed blue line: $\gamma_2 = \chi_1 =0$, $\chi_2=-2 \chi$. 
The optimization over the interferometer directions is done 
independently in the $(N_0-2)$ subspace.
Insets: Husimi functions for the  values of $\gamma_i$ and $\chi_i$
corresponding to the circles on the curves. Other parameters as in Fig.~\ref{fig2}.
}
\end{figure}

Figure \ref{fig3} shows the Fisher information $F_{N}(t_2)$ in the subspace with  $N=N_0-2$ atoms, 
corresponding to the state conditioned to a single loss event.
For symmetric energies, the
low values of $F_N$ at small symmetric loss rates are direct
consequences of the cancellation among channels (see Appendix~\ref{sec-app3}).  
For $\gamma_2=0$ much larger values are found, showing that the 
aforementioned non-vanishing inter-component coherences carry useful quantum correlations. 
 At intermediate rates, $F_N(t_2)$ increases  
and reaches a maximum  as a result of the reduced 
phase noise when $\gamma_1=\gamma_2$
(note that this peak will not be seen on $F_{\rm tot} (t_2)$ because of 
the rapid decay of the probability $w_{N_0{-}2}(t)$ by increasing $\gamma_m$). 
This reduction is clearly seen on the Husimi
distributions $ Q_{N} (\theta,\phi) = \frac{1}{\pi}\bra{N; \theta ,
  \phi}\hat{\rho}_N (t_2)  \ket{N; \theta , \phi}$ 
which display a flat profile for small losses and
two emerging peaks for larger losses (insets in Fig.\ \ref{fig3}). 
For asymmetric losses $\gamma_2=0$, by choosing $\chi_1=0$ we have $F_N(t_2) \rightarrow N^2$ 
in the small loss limit and the Husimi function has two peaks at the CS phases $\phi=\pm \pi/2$, 
in agreement with our prediction that the conditional state 
converges to a two-component superposition; at intermediate rates $F_N(t_2)$ decreases with $\gamma_1$
because the Gaussian damping compensates phase noise reduction.  
At large losses this damping  transforms $\hat{\rho}_{N_0-2}(t_2)$ into a superposition
of Fock states with $n_1=0$ or $1$ atoms in the first mode~\cite{long_article}. 
For symmetric losses and  even $N_0$ , $\hat{\rho}_{N_0-2}(t_2)$
 is transformed instead into 
the Fock state $\ket{\frac{N}{2}, \frac{N}{2}}$  
having a larger Fisher information $\sim N^2/2$. 

\begin{figure}
\centering
\includegraphics[width=0.92\columnwidth]{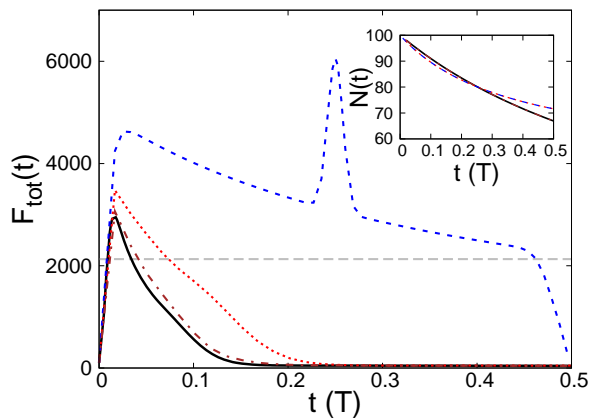}
\caption{\label{fig1} (Color online) Total quantum Fisher
information $F_{\rm tot} (t)$ vs 
time $t$ (in units of  $T$).
 From top  to bottom: $ \gamma_1 T=8/300$, $\gamma_2 = \chi_1 =0$, 
$\chi_2  = - 2 \chi$ (blue dashed line); 
$\gamma_1 T= 8/300$, $\gamma_2=0$, $\chi_1  = - \chi_2  = \chi$ (red
dotted line);  
$\gamma_1 T = \gamma_2 T = 1/100$, $\chi_1=0$, $\chi_2 = - 2 \chi$ (brown 
dot-dashed line);
$\gamma_1 T = \gamma_2 T =1/100$, $\chi_1  = - \chi_2  =\chi$ (black solid
line).
Other parameters as in Fig.~\ref{fig2}.
Dashed horizontal line: $F_{\rm tot}$ for the highest squeezed
state in the lossless case. 
Inset: average number of atoms vs time for the same parameters.
}
\end{figure}

Our main result is presented in Fig.\ \ref{fig1}, which displays the total Fisher information
$F_{\rm tot}(t)$ obtained from an exact diagonalization
of Eq.~(\ref{eq-master_equation}) for various rates and energies, 
keeping  the same  value of $\chi$ and of the mean number of atoms at time $t_2$ (see the inset). 
We find that for $\gamma_2=0$, tuning the energies has a 
strong effect on $F_{\rm tot}(t)$ at times $\chi t \gtrsim N_0^{-1/2}$, yielding  
to larger Fisher informations than for squeezed states. 
Quantum correlations in the total density matrix $\hat{\rho}(t)$ are then preserved
even after the loss of $20\%$ of atoms.

\section{Conclusions}

We have shown that the interplay of atom losses and interactions  in BJJs leads
to different decoherence scenarios even after a single loss event. 
In particular, for strongly asymmetric two-body losses 
one can protect superpositions  by tuning  the
interaction energies, leading to useful states for high-precision atom interferometry applications. 
Such asymmetric losses occur in internal BJJs 
with $^{87}$Rb atoms in the hyperfine states $|F, m_F\rangle = |2, -1\rangle$ and $|1, 1\rangle$, 
subject respectively to fast
two-body  and  slow one-body losses~\cite{gross2010dissJPB}.
Finally, we note that the $\phi$ noise associated to the loss times $s_K$
could be used to obtain an indirect measurement of $s_K$  from the interferometric estimation
of the phase rotation of the MSCSs, 
yielding to an observation of  quantum jumps like in cavity QED~\cite{Haroche}.

\acknowledgments
We are grateful to K.~Rz\k a\.zewski,
F.~Hekking, P.~Treutlein, C.~Gross, and M.~Oberthaler for 
helpful discussions. D.S and K.P acknowledge support from the project 
ANR-09-BLAN-0098-01, 
A.M. from the ERC Handy-Q project, and K.P. from the Polish Government Funds for the years
2010-2012
and the project ``Decoherence in long range interacting quantum systems and devices'' 
sponsored by the Baden-W\"urttemberg Stiftung.  

\appendix
\section{Derivation of Eq.~(\ref{eq-wave function_one_jump_bis_bis})} \label{sec-app1} 

Let us show that the quantum trajectory $t \mapsto \ket{{\psi}_{1} (t)}$ 
having one loss event in channel $m$ at the random time $s \in [0,t]$ is given by
Eq.~(\ref{eq-wave function_one_jump_bis_bis}).
We prove this formula for $m=1$ (the other cases $m=2$ and $m=12$ are similar). 
We first determine how  an initial Fock state
$\ket{n_1,n_2}$ is transformed if two atoms  are lost in mode $1$ at time $s$.
Using Eq.~(\ref {eq-no_jump}), this state becomes  
 $\sqrt{n_1(n_1-1)}  e^{-\I \Phi_{t,s}(n_1,n_2)}  \ket{n_1-2,n_2}$ at time $t$, where 
$\Phi_{t,s} (n_1,n_2)= (t -s) H_{\rm eff} (n_1-2,n_2) + s  H_{\rm eff} ( n_1,n_2)$ is a 
 complex dynamical phase and  $H_{\rm eff} (n_1,n_2)$ are the quadratic eigenvalues  of $\Heff$. 
Setting $n_2=N_0-n_1$ yields
\begin{eqnarray} \label{eq-A1}
\nn
\Phi_{t,s} (n_1,n_2) 
& = &  t H_{\rm eff} (n_1-2,n_2) +
n_1  \phi_{1} (s) 
\\
& & + \I n_1 \ln \Bigl( \tan \Bigl( \frac{\theta_{1} (s)}{2} \Bigr) \Bigr)
\end{eqnarray}
up to $n_1$-independent constants,
with $\theta_1 (s)  =  2 \arctan ( e^{- s (2 \gamma_1-\gamma_{12})} )$ and
$\phi_1 (s)  =  2 s ( U_1 - U_{12} )$.  
The two last terms in (\ref{eq-A1}) correspond 
respectively to  the energy and damping changes due to the atom loss at time $s$.
Replacing $\ket{n_1,n_2}$ in the Fock-state expansion
of the initial CS [see Eq.~(\ref{eq-def_coherent_state})] 
by the above transformed state, we get Eq.~(\ref{eq-wave function_one_jump_bis_bis}).

\section{Density matrix in the subspace with $(N_0-2J)$ atoms} \label{sec-app2} 

We now determine the conditional density matrix $\hat{\rho}_{N_0-2}(t)$ for a single loss event ($J=1$)
in the Fock basis.
Using Eq.~(\ref{eq-no_jump}) we find the distribution
\begin{eqnarray}
\non
& & p_m^{(t)} (s;1)= \gamma_m e^{-s G_m} [\cosh ( s\delta_{m} )]^{N_0-2} 
\\
\non
& & \hspace*{5mm}  \times \| e^{-i t \Heff} \ket{N_0-2;\theta_m(s),0}\|^2 \frac{N_0 ( N_0-1)}{4}
\end{eqnarray}
with
$G_{i}=(2 \gamma_i + \gamma_{12}) N_0 -2 \gamma_{i} -2 \gamma_{12}$, $i=1,2$, and
$G_{12} = (\gamma_1 + \gamma_2 + \gamma_{12}) N_0 - 2 \gamma_1 - 2 \gamma_2 - \gamma_{12}$.
Averaging over trajectories as in Eq.~(\ref{eq-density_matrix_subspace_N_0-j}) 
yields
\begin{eqnarray} \label{eq-density_matrix_N_0-2_sector_in_Fock_basis}
& &  \bra{n_1 ,n_2}  \hat{\rho}_{N_0-2}(t ) \ket{n_1',n_2'} 
\\
\non
& & \hspace*{5mm} 
 \propto  
   \sum_m \gamma_m  C_m (t; n_1, n_1')  \bra{n_1 ,n_2} 
\widetilde{\rho}_{N_0-2}^{\rm{(no\,loss)}} (t ) \ket{n_1',n_2'} ,
\end{eqnarray}
where
$\widetilde{\rho}_{N_0-2}^{\rm{(no\,loss)}} (t) 
 =  e^{-\I t \Heff} \ketbra{N_0\!-\!2;\frac{\pi}{2},0}{N_0\!-\!2;\frac{\pi}{2},0}  e^{\I t
\Heff^\dagger}$ is the conditional state having no loss in
$[0,t]$ for an  
initial CS with $N_0-2$ atoms, and
\begin{equation}
\non 
C_m ( t; n,n^\prime )\! =\! \frac{1-e^{-t [G_m + \delta_m (n+n'-N_0+2) + 2 \I \chi_m
(n-n') ]}}{G_m\!+\! \delta_m (n+n'\!-\!N_0\!+\!2) \! +\! 2 \I \chi_m (n-n')}.
\end{equation}
Equation~(\ref{eq-density_matrix_N_0-2_sector_in_Fock_basis}) shows  that the density matrix conditioned 
to a single loss event is a superposition of CSs with $N_0-2$ atoms modulated  by 
the envelope $\sum \gamma_m C_m ( t; n,n' )$ and the  damping factor of Eq.~(\ref{eq-state_no_jump}). 

The above calculation can be generalized to trajectories with $J>1$
jumps between $0$ and $t$.
For  $\gamma_m t\ll 1$ and  $J \ll N_0$, 
$\hat{\rho}_{N_0-2J} (t )$ is still given by
Eq.~(\ref{eq-density_matrix_N_0-2_sector_in_Fock_basis}) upon replacing
 $\sum \gamma_m C_m(t;n_1,n_1')$ by 
 $[ \sum_{m}  \gamma_{m} C_{m} (t; n_1, n_1') ]^J$.

\section{Weak loss regime} \label{sec-app3} 

In this appendix we study the density matrix (\ref{eq-density_matrix_N_0-2_sector_in_Fock_basis})
in the subspace with $N=N_0-2$ atoms in 
 the weak loss regime  $\gamma_m \ll q \chi/N_0$. 

One can show~\cite{ferrini2010} that the density matrix $\hat{\rho}^{(0)}(t_q)$
associated with the superposition of CSs $\sum_k c_k \ket{N;\frac{\pi}{2},\phi_k}$ 
formed at time $t_q$ in the absence of losses has the following structure in the Fock basis:
$\langle {n_1,n_2} |\hat{\rho}^{(0)}(t_q) | {n_1',n_2'} \rangle$ 
is the sum of two matrices, the first one having 
zero matrix elements for $n_1' \not= n_1$ modulo $q$ and corresponding to the
statistical mixture of CSs 
$\sum_{k} |c_k|^2  \ketbra{N;\frac{\pi}{2},\phi_k}{N;\frac{\pi}{2},\phi_k}$, 
and the second
one having zero  matrix elements for $n_1' = n_1$ modulo $q$ and corresponding to the coherences
$\sum_{k\not=k'} c_k c_{k'}^\ast  | {N;\frac{\pi}{2},\phi_k} \rangle \langle {N;\frac{\pi}{2},\phi_{k'}}|$.

For weak losses  $\gamma_m \ll q \chi/N_0$ and equal interaction energies $U_1=U_2$,
\begin{equation}
|C_m (t_q;n,n')|\simeq \frac{\pi}{|\chi| q} \Bigl| \sinc ( \frac{\pi (n-n')}{q} ) \Bigr|
\end{equation}
vanishes at the values $n'=n+pq$ ($p=\pm 1,\pm 2,\cdots$)  corresponding to the nonzero
off-diagonal elements of the aforementioned statistical mixture of CSs,
and decays like $|n'-n|^{-1}$ for the other off-diagonal elements encoding the coherences. 
Hence the statistical mixture of CSs is transformed after one loss event into a mixture of Fock states
(complete phase relaxation). 
The inter-component coherences are, however, non vanishing,
except for $\gamma_1=\gamma_2$ and $q=2$, because
then the contributions of the two channels cancel each other,
 $\sum_m\gamma_m C_m ( t_2; n,n' ) \simeq 0 $ for $n\neq n'$.
This explains  the low and high values of $F_{N_0-2}(t_2)$ for 
symmetric and asymmetric losses seen in Fig.\ 2 (recall that
the coherences are responsible for the quantum correlations quantified by the Fisher information). 

Taking now asymmetric interaction energies, $\chi_1=0$ and $\chi_2=-2\chi$,  
one infers from
\begin{equation}
\sum_m\gamma_m C_m ( t_2; n,n' ) \simeq \frac{\pi}{2\chi} \bigl( \gamma_1 +  \gamma_2 \delta_{n n'} \bigr)
\end{equation}
that  $\hat{\rho}_{N_0-2}(t_2 ) \propto \widetilde{\rho}_{N_0-2}^{\rm{(no\,loss)}} (t_2 )$ when $\gamma_2=0$.
Therefore, the conditional state is a two-component superposition slightly modified by Gaussian damping.


\end{document}